\documentclass[pra,twocolumn,superscriptaddress,showpacs,floatfix]{revtex4-1}

\usepackage{epsfig,graphicx,amsmath,amsfonts,amssymb,times}

\usepackage[sort&compress]{natbib}
\begin{document}
	
%\title{Mesoscopic decoherence-free subspaces and metastable pointer states in correlated coherent states under dissipation }
\title{Metastable decoherence-free subspace and pointer states in mesoscopic quantum systems}

\date{\today}

\begin{abstract}
Two initially correlated coherent states, each interacting with its own independent dissipative environment exhibit a sudden transition from classical to quantum decoherence.  This change in the dynamics is a turning point in the decoherence, in the sense that depending on the average number of photons of each cavity, decoherence can even be suppressed. Indeed, the quantum state is time-independent for a time span in the mesoscopic regime, revealing a decoherence-free subspace.  Furthermore, the absence of decoherence is manifested in the apparition of a metastable pointer state basis.
\end{abstract}

\author{F. Lastra}
\affiliation{Departamento de F\'{\i}sica, Facultad de Ciencias B\'{a}sicas, Universidad de Antofagasta,   
Casilla 170,  Antofagasta, Chile}

\author{C.E. L\'opez}
\affiliation{Departamento de F\'{\i}sica, Universidad de Santiago de
Chile, USACH, Casilla 307 Correo 2 Santiago, Chile}

\author{J. C. Retamal}
\affiliation{Departamento de F\'{\i}sica, Universidad de Santiago de
Chile, USACH, Casilla 307 Correo 2 Santiago, Chile}
\affiliation{Center for the Development of Nanoscience and Nanotechnology, 9170124, Estaci\'on Central, Santiago, Chile}
\maketitle

The study of quantum correlations in multipartite quantum systems  is a central problem in quantum mechanics. The study of  all correlations existing  between two quantum systems, not only entanglement,  has captured the attention of many researchers in last years. Quantum and classical correlations  embodied in a bipartite quantum system are contained in the quantum mutual information\cite{quantummutual,oli}. From this fundamental concept, the search of all quantum correlations that can be found in a bipartite system has motivated the introduction of quantum discord \cite{oli, hen,opp,Luo,Luo1,mod}. One  main result driving research in this field is that quantum computation is possible even in the absence of quantum entanglement  \cite{lan}. 

A central issue is the understanding of how  a bipartite system behaves under the interaction with an environment. Concerning this effect, it is widely known that entanglement  could vanish suddenly depending on the initial state \cite{zy,dio,dod,yu,san}.  Given that, it is important to know how quantum correlations, other than entanglement,  are affected by the presence of environment \cite{sha,dat,pia,wer,maz1}. Both classical and quantum correlations are affected such that in most cases they decay asymptotically independently on the initial state, this feature distinguishes them from quantum entanglement. As the total mutual information decays as a function of time, it could be interesting to  see how this decay can be associated to classical or to quantum correlations.   An unexpected  behavior has been revealed for a certain class of quantum states evolving under dephasing: Their classical correlations decay while the quantum correlations remain constant. This is followed by an exchange in the roles, that is, decaying of quantum correlations and freezing of classical correlations \cite{maz, mazz}. These findings have been  experimentally observed for non-markovian and markovian reservoirs \cite{xu,cor}.  Freezing of classical correlations has been shown in non dissipative decoherence dynamics \cite{Chanda15,lastra17}. Moreover, such behavior reveals the apparition of a pointer basis \cite{oli,hen,cor,lastra2014,lastra17,zur}.

On the other hand, coherent states and environment effects on its quantum coherence have been one of the most important problems since the beginning of  quantum mechanics. A distinctive feature concerning its dynamical behavior is the appearance of a decoherence time scale depending on the distance between  coherent states, which is much  shorter than the decay of any other observables~\cite{Haroche96,Haroche96PRL}.  Moreover, in last years these states have proven to be useful in practical applications such as quantum metrology~\cite{Haroche16,Nolan17,Viola07}.  In this manuscript, we address the evolution of quantum and classical correlations of an initially incoherent superposition of entangled coherent states. We find different time scales that give rise to a wide variety of behaviors, in particular, we observe sudden transitions in the decoherence dynamics. Moreover, measuring on one of the parties (to calculate classical correlations) projects the other into a basis which is not affected by decoherence. This reveals the apparition of  metastable pointer states and a decoherence-free subspace whose time span depends on the amplitude of the initial coherent states.

To find such features, let us first consider the problem of a single cavity mode coupled to a dissipative reservoir. In the interaction picture, the Hamiltonian of this system is given by
\begin{equation}
\hat{H}_I =  \hbarÊ\sum_k g_k \left(\hat{a} ^\dagger \hat{b}_k e^{i(\nu-\nu_k)t}+\hat{a} \hat{b}_k^\dagger e^{-i(\nu-\nu_k)t} \right)
\label{}
\end{equation}
If the cavity mode is prepared initially in a coherent state $|\alpha\rangle$ while all reservoir modes are in the vacuum state $\prod_k |0_k\rangle $, the evolution of the cavity-reservoir system will take the form:
\begin{equation}
e^{-i \hat{H}_I t/\hbar}|\alpha\rangle \prod_k |0_k\rangle=\mid \alpha e^{-\gamma t/2} \rangle \prod_{k} \mid \alpha_k \rangle
\end{equation}
with $\gamma$ the decay rate of the cavity mode. The state $|\alpha_k \rangle$ denotes a coherent state for the $k-$th mode of the reservoir with amplitude $\alpha_k =f_k \alpha$. In the Markov approximation, the factors $f_k$ satisfy the relation $\sum_k f_k^{2}=1-e^{-\gamma t}$ \cite{lastra}.

Consider now two modes  ($a$ and $b$), embedded in two noninteracting cavities affected by two independent dissipative reservoirs. Let us assume both cavities are prepared initially in an incoherent superposition such as:
\begin{equation}
\hat{\rho}_{ab} (0)=p| \psi_0^+\rangle \langle\psi_0^+\mid+(1-p)\mid \phi_0^+\rangle \langle \phi_0^+\mid
\label{ini2}
\end{equation}
where,
\begin{eqnarray}
| \psi^\pm _0\rangle &=& \frac{1}{\Lambda_\pm }\left(| \alpha \rangle_a  | \alpha \rangle_b \pm | -\alpha \rangle_a  |-\alpha \rangle_b \right)\label{psi} \\
| \phi^\pm _0\rangle &=& \frac{1}{\Lambda_\pm }\left(| \alpha \rangle_a  |-\alpha \rangle_b \pm | -\alpha \rangle_a  | \alpha \rangle_b \right)\label{phi}
\end{eqnarray}
and the normalization factor is given by $\Lambda_{\pm}^2 = 2(1\pm e^{-4\bar{n}})$.
 
In what follows, we deal with the calculation of quantum and classical correlations of the system described as a function of time. In general this is a difficult task since no closed formula exists for arbitrary states. However, for particular states such as  $X$-states \cite{xstate,CHOh2011} analytical calculations can be carried out. After some calculations is not difficult to realize that the temporal evolution for the state (\ref{ini2}) can be written as an $X$-states in the effective two-qubit basis: $\{|\eta_+\rangle_a|\eta_+\rangle_b,|\eta_+\rangle_a|\eta_-\rangle_b,|\eta_-\rangle_a|\eta_+\rangle_b,|\eta_-\rangle_a|\eta_-\rangle_b\}$, where states $|\eta_+\rangle$$(|\eta_-\rangle)$ are commonly known as Schr\"odinger cat states \cite{Haroche96,Haroche96PRL}:
\begin{equation}
|\eta_{\pm} \rangle=\frac{1}{\Gamma_{\pm}}(\mid \alpha_t \rangle \pm \mid -\alpha_t \rangle),
\label{entan1}
\end{equation}
with $\alpha_t^2 = \bar{n} \exp{(-\gamma t)}$. In such basis we obtain the following $X$-state 
\begin{equation}
\begin{tabular}{l}
$\rho _{ab} (t)=\left(
\begin{array}{cccc}
r_{11} & 0 & 0 & r_{14} \\0 & r_{22} & r_{23} & 0 \\
0 & r_{32} & r_{33} & 0 \\
r_{41} & 0 & 0 & r_{44}%
\end{array}%
\right).$%
\end{tabular}
\label{Xstate}
\end{equation}%
where the matrix elements are:
\begin{eqnarray}
r_{11} &=& \frac{1}{16} \left[\frac{\Gamma_+(t) \Gamma_+(t)}{\Lambda_+ } \bar{\Lambda}_+ (t) \right]^2, \notag \\
r_{22} &=& \frac{1}{16} \left[\frac{\Gamma_+ (t)\Gamma_- (t)}{\Lambda_+ } \bar{\Lambda}_- (t) \right]^2 = r_{33}, \notag\\
r_{44} &=& \frac{1}{16} \left[\frac{\Gamma_-(t) \Gamma_-(t) }{\Lambda_+ } \bar{\Lambda}_+ (t) \right]^2, \label{pop}\\
r_{14} &=& r_{41} = \frac{1}{16} \left(2p - 1\right) \left[\frac{\Gamma_+(t) \Gamma_-(t)}{\Lambda_+ } \bar{\Lambda}_+ (t) \right]^2, \notag\\
r_{23} &=& r_{32} = \frac{1}{16} \left( 2p-1\right)\left[\frac{\Gamma_+(t) \Gamma_-(t)}{\Lambda_+ } \bar{\Lambda}_- (t) \right]^2.\notag 
\label{coeficientes}
\end{eqnarray}
with $\Gamma_{\pm}^2 = 2(1\pm \exp{(-2\alpha_t ^2)})$, $\bar{\Lambda}_{\pm}^2(t) = 2(1\pm e^{-4\bar{\alpha}_t ^2})$ and $\bar{\alpha}_t^2 = \bar{n} \left(1- e^{-\gamma t}\right)$.  

A closer look on these analytical expressions, can shed some light on relevant times scales. As will be shown below, the dynamics exhibits different behaviors associated to these relevant times. The evolution of matrix elements in Eq. (\ref{pop}) are mainly determined by the terms $\alpha_t^2 $ and $\bar{\alpha}_t^2$ in the exponentials. On one hand, for small values of $\bar{n}$  $(\bar{n} \lesssim1 )$, as time goes by we see that the value of $\alpha_t$ decreases leading to a slower decay of terms involving $e^{-\alpha_t ^2}$. On the other hand, the value $\bar{\alpha}_t^2$  increases, leading to a faster decay of terms involving $e^{-\bar{\alpha}_t ^2}$. However, in the mesoscopic limit $\bar{n} \gg 1$, at short times the term $\alpha_t$ is large enough so that $e^{-\alpha_t^2 }\simeq0$ and then the dynamics is dominated by  the term $e^{-\bar{\alpha}_t^2 }$. This means that, the time dependence in Eq.(\ref{pop}) are only in the terms $\bar{\Lambda}_\pm (t)$ while $\Gamma_\pm (t)$ are constant. On the other hand when $\gamma t \rightarrow \infty$ we have the opposite case: the dynamics is now governed by $\Gamma_\pm (t)$ while $\bar{\Lambda}_\pm (t)$ are constant. For intermediate times  both $\alpha_t$ and $\bar{\alpha}_t$ are large enough so that  $\bar{\Lambda}_\pm (t)$ and $\Gamma_\pm (t)$ are both constant, so that the matrix elements  given by:
\begin{eqnarray}
r_{11} &=& r_{22}  = r_{33}  = r_{44}  = \frac{1}{4} \label{coef1}\\
r_{14} &=& r_{41}  = r_{23}  = r_{32}  =\frac{1}{4}\left(2p-1\right),\label{coef2}
\end{eqnarray}
exhibit no evolution and depend only on the initial condition. 
\begin{figure}[t]
\includegraphics[width=90mm]{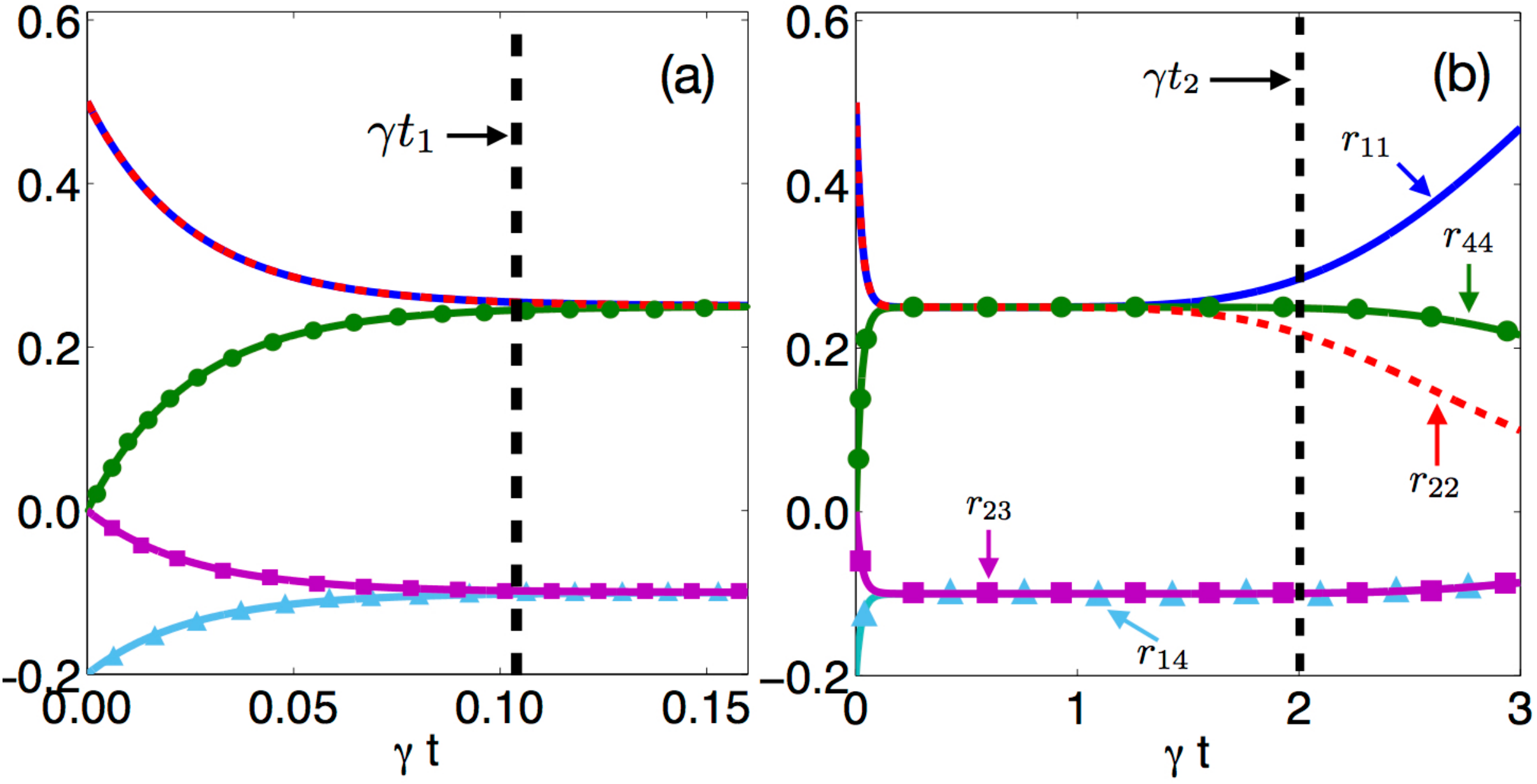}
\caption{Evolution of the density matrix elements of Eq.~(\ref{pop}) for $\bar{n} = 10$. Times $t_1$ and $t_2$ are in this case: $\gamma t_1 \simeq 0.1$ and $\gamma t_2 \simeq 2$ and are shown as vertical dashed lines.}  
\label{figpop}
\end{figure}

We can estimate the value of  times that define this dynamical regime. For large $\bar{n}$, the exponentials are relevant if $\alpha_t \lesssim 1$ or $\bar{\alpha}_t \lesssim 1$, respectively. This allows us to define the times $t_1$ from $\bar{\alpha}_{t_1} =1$ and $t_2$  from $\alpha_{t_2} = 1$, such that:
\begin{eqnarray}
t_1 &=& \frac{1}{\gamma}\ln{\frac{\bar{n}}{\bar{n}-1} } \notag \\
t_2 &=& \frac{1}{\gamma}\ln{\bar{n}} 
\label{t1t2}
\end{eqnarray}
The evolution of the matrix elements in Eq. (\ref{pop}) is shown in Fig. \ref{figpop}, for $\bar{n} = 10$. Interestingly, between $t_1$ and $t_2$ we observe that the system seems to be unaffected by decoherence, that is, it settles on a { \it metastable decoherence-free subspace} evidenced by Eqs. (\ref{coef1}) and (\ref{coef2}). As we will address further in this manuscript, these results are crucial in the quantum and classical correlations dynamics.

We can now focus on the study of quantum and classical correlations. A bipartite quantum system $\hat{\rho}_{{ab}}$ as the one described above, can feature both quantum and classical correlations. Total correlations are characterized by the quantum mutual information $I(\hat{\rho}_{{ab}})=S(\hat{\rho}_{{a}})+S(\hat{\rho}_{{b}})-S(\hat{\rho}_{{ab}})$,
where $S(\hat{\rho})=-{\rm Tr}[\hat{\rho}\log_2(\hat{\rho})]$ is the
von Neumann entropy. Based on this expression
correlations can
be separated according to their classical and quantum
nature, respectively. In this way the quantum discord
has been introduced as $D(\hat{\rho}_{ab})=I(\hat{\rho}_{ab})-C(\hat{\rho}_{ab})$, where $C(\hat{\rho}_{ab})$ are the classical correlations defined by 
\begin{equation}
C(\hat{\rho}_{ab})=\max_{\{\hat{\Pi}_{k}\}}\left[S(\hat{\rho}_{a})-S(\hat{\rho}_{ab}\mid\{\hat{\Pi}_{k}\})\right],\label{cc}
\end{equation}
here, the optimization is carried out with respect  all possible complete set of projector
operators  $\{\hat{\Pi}_{k}\}$ for the subsystem
$b$, and $S(\hat{\rho}_{ab}\mid\{\hat{\Pi}_{k}\})=\sum_{k}p_{k}S(\hat{\rho}_{k})$,
$p_{k}={\rm Tr}(\hat{\rho}_{ab}\hat{\Pi}_{k})$, and $\hat{\rho}_{k}={\rm Tr}_{b}(\hat{\Pi}_{k}\hat{\rho}_{ab}\hat{\Pi}_{k})/p_{k}$. In general, the   optimization is a difficult problem to address, however, for states of the form of Eq.~(\ref{Xstate}), classical and quantum correlations can be solved analytically \cite{xstate,CHOh2011}. Specifically, it has been shown in ref.~\cite{CHOh2011}, that the optimal observables for a real $X$-state such as Eq.~(\ref{Xstate}), corresponds to $\sigma_z$ if
\begin{equation}
\left(|r_{23}|+|r_{14}| \right)^2 \leqslant \left(r_{11}-r_{22}\right)\left(r_{44}-r_{33}\right) \label{condsz}
\end{equation}
and $\sigma_x$ if
\begin{equation}
|\sqrt{r_{11} r_{44}}-\sqrt{r_{22} r_{33}}| \leqslant |r_{23}|+|r_{14}| \label{condsx}
\end{equation}
Under these conditions, expressions for discord in \cite{xstate} are equivalent. In such case, the expression for the classical correlations are now given by

\begin{equation}
C(\hat{\rho}_{ab})=S(\hat{\rho}_{a})- \min_{\{\sigma_x,\sigma_y\}}\left[S(\hat{\rho}_{ab})\rvert \{\sigma_x,\sigma_y\}\right],\label{cca}
\end{equation}
where $S(\hat{\rho}_{ab})\rvert \{\sigma_x,\sigma_y\}$ is the von Neumann entropy of $\hat{\rho}_{ab}$ when $\sigma_x$ or $\sigma_z$ has been measured in the subsystem $b$. When the minimum occurs for $\sigma_x$, we denote the classical correlations as $C(\hat{\rho_{ab}})|_{\sigma_x}$, while if the minimum is achieved by measuring $\sigma_z$ then we denote it as $C(\hat{\rho_{ab}})|_{\sigma_z}$ instead.

\begin{figure}[t]
\includegraphics[width=85mm]{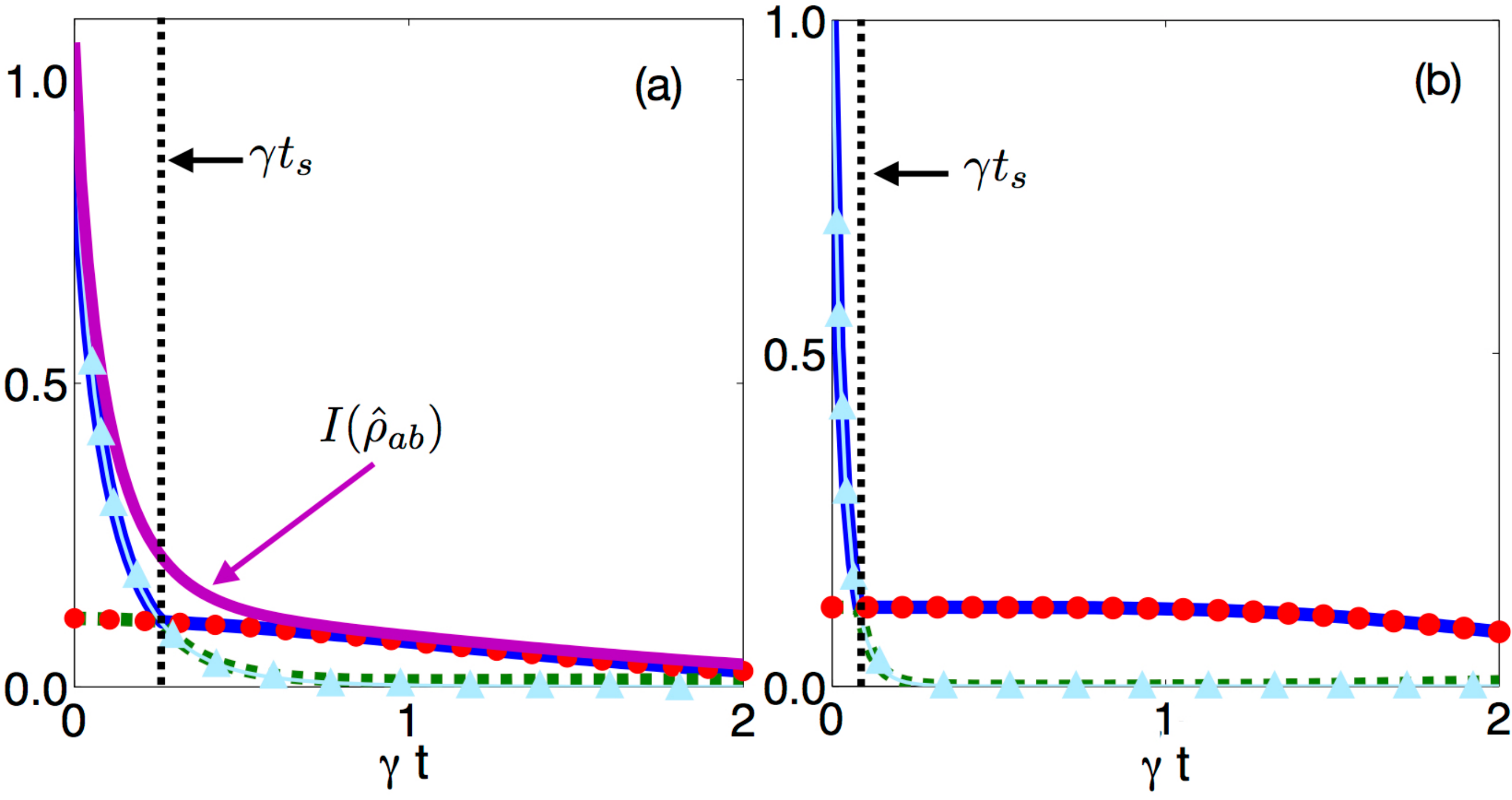}
\caption{Evolution of classical correlations (blue solid-line), discord (green dashed-line) and quantum mutual information (purple solid-line) as a function of the dimensionless time $\gamma t$  for the initial state given by (\ref{ini2}) with $p = 0.3$. $(a)$ For $\bar{n} = 1$. (b) For $\bar{n} = 3$. $C(\hat{\rho_{ab}})|_{\sigma_x}$ (red dots) and $C(\hat{\rho_{ab}})|_{\sigma_z}$ (light blue triangles) are shown in both plots.}  
\label{fig2}
\end{figure}

%\begin{figure}[t]
%\includegraphics[width=42mm]{fig1a.pdf}
%\includegraphics[width=42mm]{fig1b.pdf}
%\caption{(a) Evolution of classical correlations (red), discord (blue) and quantum mutual information (green) as a function of the dimensionless time $\gamma t$. The initial state is given by (\ref{ini2}) with $p = 0.3$ and $\bar{n} = 1$. (b) Classical correlations (red), $C(\hat{\rho_{ab}})|_{\sigma_x}$ (light blue) and $C(\hat{\rho_{ab}})|_{\sigma_z}$ (light green).}  
%\label{fig1}
%\end{figure}

Let us consider the initial state (\ref{ini2}), corresponding to an incoherent superposition of entangled coherent states.
In Fig.~\ref{fig2}, we show the evolution of correlations in the two-mode subsystem as a function of $\gamma t$ for $(a)$ $\bar{n}=1$ and for (b) $\bar{n}=3$. As evidenced by the quantum mutual information (purple solid-line) in Fig.~\ref{fig2}(a), the overall system correlations decays smoothly as a consequence of decoherence. However, we can observe a sudden change in the dynamics of classical correlations and also in quantum discord at $t=t_s$. This tells us that there is a sudden change in the decoherence dynamics: Before $t_s$ decoherence has mostly a classical component (classical correlations decay faster than discord); after $t_s$ the roles are inverted and decoherence has mostly a quantum component. Despite this, decoherence has still mixed classical and quantum contributions which is different to what has been found for example in ref.~\cite{mazz}, where the decoherence has either a quantum character or a classical one, but never both. On the other hand, Fig.~\ref{fig2}(b) shows that by increasing the mean photon number $\bar{n}$ of the cavity modes, the classical correlations nearly freezes between $\gamma t_s$ and $\gamma t\approx 1$.

%\begin{figure}[t]
%\includegraphics[width=42mm]{fig2a.pdf}
%\includegraphics[width=42mm]{fig2b.pdf}
%\caption{(a) Evolution of classical correlations (red), discord (blue) and quantum mutual information (green) as a function of the dimensionless time $\gamma t$. The initial state is given by (\ref{ini2}) with $p = 0.3$ and $\bar{n}=3$. (b) Classical correlations (red), $C(\hat{\rho_{ab}})|_{\sigma_x}$ (light blue) and $C(\hat{\rho_{ab}})|_{\sigma_z}$ (light green).}  
%\label{fig2}
%\end{figure}

\begin{figure}[t]
\includegraphics[width=87mm]{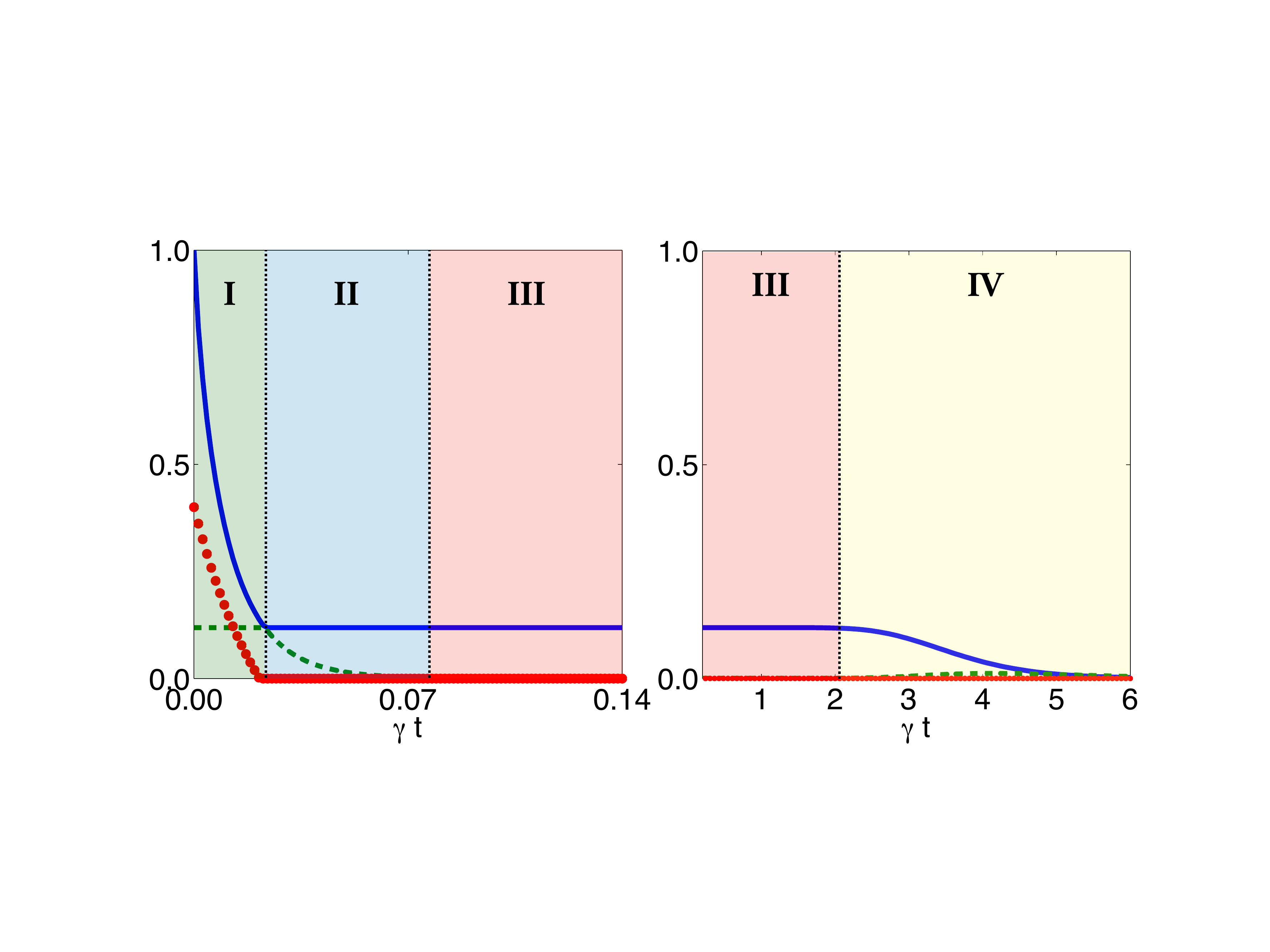}
\caption{Evolution of classical correlations (blue solid-line), discord (green dashed-line) and concurrence (red dots)  as a function of the dimensionless time $\gamma t$. The initial state is given by (\ref{ini2}) with $p = 0.3$ and $\bar{n} = 10$. Left plot is for short times $\gamma t <1$  and the right plot is for larger times.}  
\label{figstrong}
\end{figure}

As we previously mentioned, when exploring the mesoscopic limit of a large number of photons in the cavity, interesting features in the dynamics are revealed.  Such behavior is extended into quantum and classical correlations: This is clear in Fig.~\ref{figstrong} where the dynamics of correlations is shown for $\bar{n} = 10$. In such figure, four dynamical regimes can be identified: Regime (I) where classical correlations decay as a result of the decoherence process while discord is constant. Regime (II) is  determined by a decaying discord and  a frozen classical correlations. In regime (III) both discord and classical correlations attain a constant value, in particular discord vanished. Finally, in regime (IV) classical correlations start to decay again while discord shows a revival and then decays asymptotically to zero as all the energy in the cavities is transferred to the reservoirs. 

The first regime (I) of the dynamics can be interpreted as follows: the decoherence has only a classical contribution since quantum correlations are frozen. Now, at the second regime (II), we found that the decoherence process has now only a quantum contribution.  That is, regimes (I) and (II) are separated by a sudden transition from classical to quantum decoherence.  So far, the dynamics of correlations resembles the results found in \cite{mazz,lastra2014} for the case of qubits under the onset of dephasing. 

Also, Fig.~\ref{figstrong} shows that the time $t_s$ when the sudden transition from classical to quantum decoherence occurs, decreases with $\bar{n}$. This transition time besides depending on $\bar{n}$, also depends on the parameter $p$ of the initial mixed state Eq.~(\ref{ini2}). Using Eqs.~(\ref{condsz}) and~(\ref{condsx}) we find that this time $t_s$ is given by:

\begin{equation}
t_s=-\frac{1}{\gamma}\ln(1+\frac{1}{4\bar{n}}\ln \mid 2p-1\mid )
\end{equation}

\begin{figure}[t]
\includegraphics[width=50mm]{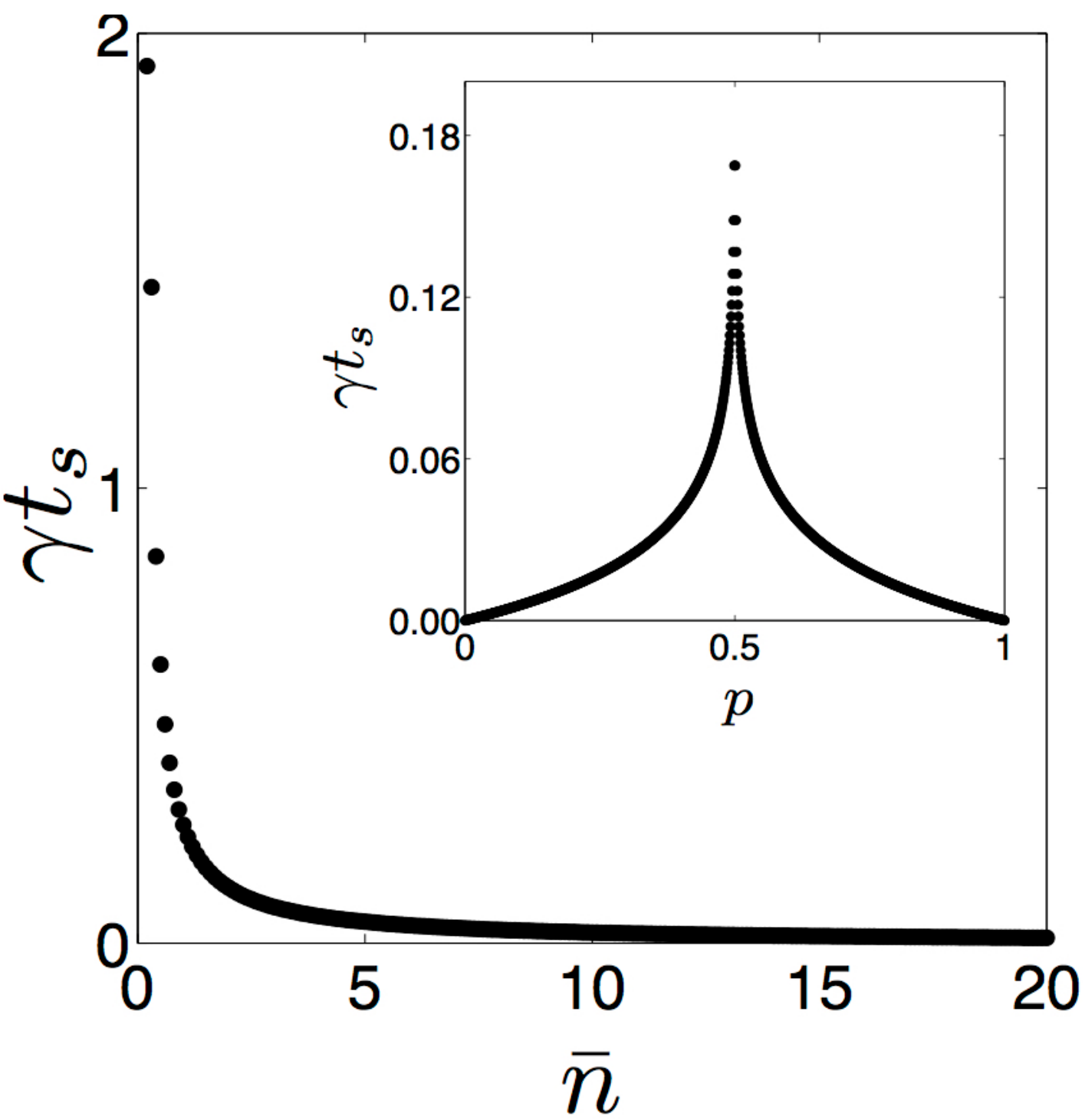}
\caption{Behavior of $t_s$ as a function of the mean photon number $\bar{n}$ for $p=0.3$ and as a function of $p$ (inset) for $\bar{n} = 10$}.   
\label{figtsrs}
\end{figure}

Fig.~\ref{figtsrs},  shows the dependence of $t_s$  on $\bar{n}$ and the initial state parameter $p$. We can observe from the figure that the time $t_s$ of the sudden transition from classical to quantum decoherence decays with $\bar{n} $ and also decays when the initial incoherent  state (\ref{ini2}) is more unbalanced, i.e, has more purity. 

On the other hand, analyzing the correlations dynamics in the strong-field case (Fig.~\ref{figstrong}) we find that in the stage III of such case, both correlations are constant, i.e., neither discord nor classical correlations are affected by decoherence. Interestingly, this is true only in a finite time span. This suggests that there is a time interval where the system settles in a time-dependent decoherence-free subspace. This  time can be estimated from Eqs. (\ref{t1t2}), leading to $\Delta t = t_2 - t_1 = (1/\gamma) \ln {\left(\bar{n}-1\right)}$. In this time interval populations are constants. Moreover, coherences are also equals and constants, that is, in this stage no decoherence is exhibited by the density matrix (\ref{Xstate}). After this finite time interval, populations $r_{22}$, $r_{33}$, $r_{44}$ and coherences $r_{14}$ and $r_{23}$ decay asymptotically to zero while $r_{11}$ goes to 1 since at $\gamma t \rightarrow \infty$ this density matrix element corresponds to the vacuum state population.
Furthermore, when classical correlations attained a constant value as in stages (II) and (III) of dynamics, it can be argued from its definition, that measurements on the second qubit (cavity mode) projects the system into a basis which is not affected by decoherence, that is, a {\it pointer state basis} \cite{cor}. However, this pointer state basis is not stable as shown in the stage (IV) in Fig.~\ref{figstrong} where classical correlations return to decay, i. e., the system settles along stages (II) and  (III) in a metastable pointer state~\cite{lastra2014,lastra17}. Notice that this occurs under a dissipative dynamics, which differs from the results showed in reference \cite{cor} where in the amplitude damping case, no pointer states are found, but only in the dephasing case. Finally in Fig.~\ref{figstrong} we have plotted entanglement using concurrence~\cite{wootters98} where we observe that entanglement suffers a sudden death~\cite{yu, Lopez08} previous to $t_s$ in Regime (I). Therefore,  along the decoherence-free time span no entanglement is present which is consistent with previous results \cite{lastra17}.

%\section{Conclusion}q

In summary the dynamics of two initially correlated  coherent states, each interacting with its own independent dissipative environment  has been analyzed. We found a sudden transition from classical to quantum decoherence since first the decoherence has only a classical component given that discord is constant. Then only discord decays meaning that decoherence is merely quantum. This sudden transition leads in the mesoscopic regime to the apparition of a metastable decoherence-free subspace. This is evidenced in the density matrix element which do not evolve during a time span. The time when this sudden transition occurs is showed to be dependent on the average number of photons in each cavity and also on the purity of the initial state. The metastable decoherence-free subspace is linked to the metastability of a pointer basis where classical correlations between cavities are frozen. On the other hand the size of the time interval depends mainly on the average number of photons in the cavities. This can be also understood in the context of pointer states: measurements on the second qubit (cavity mode) to calculate correlations, projects the system into a basis which is not affected by decoherence.

%\section{Acknowledgments }
Authors acknowledge financial support from DICYT 041631LC, Fondecyt 1140194 and Financiamiento Basal FB 0807 para Centros Cient\'ificos y Tecnol\'ogicos de Excelencia.

\end{document}